\documentclass[aps,preprintnumbers]{revtex4}
\usepackage{graphicx}

\newcommand{\bk}{{\bf k}}

\newcommand{\bP}{{\bf P}}

\newcommand{\rg}{\rm g}
\newcommand{\gb}{\bar{\rm g}}

\def\bk{{\bf k}}

\def\bX{{\bf X}}
\newcommand{\bXp}{{\bf X}^\prime}

\newcommand{\rmi}{{\rm i}}
\newcommand{\rme}{{\rm e}}
\newcommand{\rmd}{{\rm d}}
\newcommand{\w}{\omega}

\newcommand{\la}{\lambda}

\newcommand{\etab}{\bar{\eta}}
\newcommand{\etap}{\eta^{\prime}}

\newcommand{\zb}{\bar{z}}
\newcommand{\zp}{{\rm z'}}

\def\(#1){(\ref{#1})}

\begin{document}
\title{Moving Atom-Field Interaction: Quantum Motional Decoherence and Relaxation}
\author{S. Shresta}
\email{sanjiv@physics.umd.edu}
\affiliation{Department of Physics, University of Maryland, College Park, Maryland 20742}
\author{B. L. Hu}
\email{hub@physics.umd.edu}
\affiliation{Department of Physics, University of Maryland, College Park, Maryland 20742}
\preprint{umdpp 03-013}
\date{\today}
\begin{abstract}
The reduced dynamics of an atomic qubit coupled both to its own quantized center of mass motion through the spatial mode functions of the electromagnetic field, as well as the vacuum modes, is calculated in the influence functional formalism. The formalism chosen can describe the entangled non-Markovian evolution of the system with a full account of the coherent back-action of the environment on the qubit. We find a slight increase in the decoherence due to the quantized center of mass motion and give a condition on the mass and qubit resonant frequency for which the effect is important. In optically resonant alkali atom systems we find the effect is negligibly small. The framework presented here can nevertheless be used for general considerations of the coherent evolution of qubits in moving atoms in an electromagnetic field.
\end{abstract}
\maketitle

\section{Introduction}
Atomic motion is an unavoidable element in the consideration of any AMO system and an integral part of experimental designs in atom trapping devices. At issue here is the interaction between the internal degrees of freedom of an atom, assumed to contain an effective two-level system (qubit), and the electromagnetic field (EMF), modified by the atom's quantal motional degree of freedom. This problem has two aspects: 1) How does the two level activity affect the atomic motion? and 2) How does atomic motion affect the two level activity? The first aspect is the basis for laser cooling and atom trapping, which have been studied in great detail and successfully implemented by well-known experiments (for reviews see \cite{CCT1, CCTandWDP, phillips98, CCT2}). This paper is aimed at the second aspect, specifically, how quantized motion affects the qubit-EMF system dynamics, which is of interest in the design of quantum computers based on atomic qubits (in the form of a neutral atom \cite{brennen, calarco, deutsch, milburn, coldcoll} or ion \cite{poyatos}) in a QED cavity or optical potential. Effects on internal dynamics due to quantized center of mass (COM) motion have previously been studied in the situations of an atom in free space \cite{DubetskyBerman96}, in a cavity \cite{WilkensMeystre92, RenCarmichael95, VernooyKimble97}, and when the atom's qubit and COM degrees of freedom are entangled \cite{GrahamWallsZoller92}. However, all have focused on spontaneous emission rather than decoherence. The present work probes the non-Markovian regime of atom-EMF interaction, under the modest aim of explicitly computing how entanglement with quantized motion through recoil affects the decoherence and relaxation rates of an atomic qubit in free space. In order to achieve that end, we first discuss two issues of importance in computations of {\it coherent} reduced dynamics, using path integral methods.\\

\noindent {\bf The importance of including back-action}\\
It is well-known that the interaction between a two-level system (2LS, or qubit) and the EMF is the primary source of its relaxation and decoherence, while effects associated with the atom's motional degrees of freedom are usually relegated to the background. Assuming that the atom moves adiabatically limits one's consideration to those circumstances wherein the external degrees of freedom act merely as a passive parameter in the environment (here comprised of the EMF and atomic motion) of our system (the qubit), with no dynamical interplay. In technical terms, this amounts to a `test-field' approximation -- that the qubit lives in a fixed environment defined by a set of parameters, amongst them the adiabatic motion \footnote{A familiar example is a thermal bath: When characterized only by its temperature one ignores its dynamical response to the system in question.}. The test field approximation leaves out effects of changes in the environment on the system. To include the effects of the environmental variables {\it dynamically} it is essential to perform a self-consistent back-action calculation. This was done for the effect of a cavity EMF on the 2LA in Ref.~\cite{AH}.\\

\noindent {\bf Full coherence requires self-consistent treatment}\\
In tackling problems where many factors enter, it is useful to isolate one factor after another so that the remaining factors of interest to us can be simplified enough to yield some solution. For quantum coherence and entanglement such simplifications can lead to erroneous results, since phase information is lost if one artificially isolates the linking components of the complete quantum system. This brings up the necessity of self-consistency in any treatment of quantum coherence and entanglement issues. In the present case of a qubit in an EMF this requires that the fully entangled system of atomic 2LS (internal), the EMF, and the center of mass (external) degrees of freedom be treated coherently as a whole and each factor involved be allowed to evolve under the influence of the others in a self-consistent manner. This self-consistency requirement leads to non-Markovian dynamics since memory effects arise naturally and are necessary to preserve maximal coherence during the evolution \footnote{A familiar example is given by Zwanzig in his discussion of the projection operator approach: one can write down two differential equations for two interacting subsystems which make up the total system, but if one decides to focus only on one of these subsystems, its dynamics is governed by an integro-differential equation with nonlocal kernels, signifying memory effects. Note that the Markov approximation underlies many common treatments of quantum systems, such as the Fermi Golden rule, the Wigner-Weisskopf form, the Pauli master equation, to name a few. It clamps down on the dynamical interactions which may result in the violation of the consistency requirement described above, and hence could yield inadequate or erroneous results pertaining to issues of quantum coherence and entanglement in certain circumstances, such as under strong interaction, at low temperature or for a supra-ohmic environment.}. \\

\noindent{\bf Grassmannian and Coherent State Representation of Influence Functional}\\
A theoretical scheme we found satisfactory in meeting these requirements is the influence functional (IF) formalism of Feynman and Vernon \cite{FeyVer} or the related closed-time-path (CTP) effective action of Schwinger and Keldysh \cite{ctp}. The influence of the environmental variables on the system of interest is incorporated in the IF (or effective action) in such a way that the equations of motion obtained for the system will already have included the back-action of the environmental variables on the system in a dynamically self-consistent manner. This scheme has been applied to a two-level atom (2LA) interacting with an electromagnetic field (EMF) in reference \cite{ABH}. There, a first-principles derivation of the general master equations is given and applied to the study of the decoherence of a 2LA in an EMF, for the cases of a free quantum field and a cavity field in the vacuum at zero temperature. The authors found that for the standard resonant type of coupling characteristic of such systems the decoherence time is close to the relaxation time.

Here we use the influence functional method for the treatment of the  back-action of the quantum field and the quantal motion of the atom on the qubit. In Section II we compute the transition amplitude between an initial and final state using a coherent state label for the (bosonic) states of the EMF and a Grassmannian for the (fermionic) 2LS. The coherent state basis allows us to identify the Hilbert space of states with a space of coherent states. The sum over all quantum evolutions is then a sum over all paths in this space. Once the transition amplitude is computed in some sufficiently simplified form, forward and backward versions can be combined and reduced to form the reduced density matrix evolutionary operator. In Section III we calculate the evolutionary operator for the reduced density matrix when the EMF and motional degrees of freedom are integrated over. We derive an equation describing the evolution of the on and off-diagonal elements, the latter is the coherence function we seek. We end in Section IV with a discussion of our results and comments on possible further developments on this subject.

\section{The Transition Amplitude}
Our system is a 2-level atom interacting with its own center of mass (COM) motion and the EMF. We begin with a modified multi-mode Jaynes-Cummings type Hamiltonian (e.g.\ dipole and rotating wave approximation, see Appendix A of Ref\cite{ABH}),
\begin{equation}
\label{hamiltonian} H =\frac{\bP^2}{2M} +\hbar \w_o S_z +\hbar\sum_\bk [\w_\bk b_\bk^\dagger b_\bk + {\rg}_\bk(\bX) S_+ b_\bk + \gb_\bk(\bX)S_- b^\dagger_\bk].
\end{equation}
The first term in the Hamiltonian is the COM kinetic energy. The next two terms are the qubit and EMF energies, respectively. The last two terms are the interaction between the qubit, EMF, and the atom's COM degree of freedom. Note that $\bP$ and $\bX$ are both operators. Coupling of the qubit to its COM motion is through the spatial mode functions of the EMF. We shall restrict our consideration to an initial vacuum EMF at zero temperature. The result of this calculation will thus be the modification of the vacuum decoherence and relaxation rates of a qubit when the effects of quantized atomic motion are included.

The first step towards obtaining the reduced system dynamics while retaining the full system's coherence is to compute the transition amplitudes between the initial and final states which are the matrix elements of the evolution operator of the full system. We do this with coherent state path integrals. For the EMF we use a bosonic coherent state representation and for the 2-level system (qubit) degree of freedom we use the Grassmannian coherent states \cite{ohnuki, perelomov}. Coherent states are by definition generated by the exponentiated operation of the creation operator and a suitable label on a chosen fiducial state:
\begin{eqnarray}
\label{emcoherentstates} |z_\bk \rangle &=& \exp(z_\bk b_\bk^\dagger) |0_\bk \rangle \\
\label{gmcoherentstates} |\eta \rangle &=& \exp(\eta S_+) |0\rangle
\end{eqnarray}
In the case of bosonic coherent states defined in Eq.~(\ref{emcoherentstates}) the label, $z_\bk$, is a complex number, and in the case of the Grassmann coherent states defined in Eq.~(\ref{gmcoherentstates}) the label, $\eta$, is an anti-commuting number. The chosen fiducial states are the EMF vacuum and the lower 2-level state, respectively.

In order for any set of states to be useful for the decomposition of the transition matrix they must have a resolution of unity. The EMF and Grassmannian coherent states have the following decompositions of unity
\begin{eqnarray}
\label{completeness}1 = \int {\rmd}\mu(z_\bk) |z_\bk\rangle\langle\zb_\bk| =\int {\rmd}\mu(\eta) |\eta\rangle\langle\etab|
\end{eqnarray}
with the measures
\begin{eqnarray}
{\rmd}\mu(z_\bk) &=& \exp(-\zb_\bk z_\bk) \nonumber \\
{\rmd}\mu(\eta) &=& \exp(-\etab\eta) \nonumber
\end{eqnarray}
Grassmann coherent states also share other well known properties of coherent states such as being non-orthogonal and eigenstates of the annihilator:
\begin{eqnarray}
\label{non-orthogonality}\langle \zb_\bk|\zp_\bk\rangle &=& \exp(\zb_\bk \zp_\bk) \mbox{ }\mbox{ }\mbox{ }\mbox{ }\mbox{ }\langle \etab|\etap\rangle = \exp(\etab \etap) \nonumber \\
\label{annihilatoreigen}b_\bk |z_\bk\rangle &=& z_\bk |z_\bk\rangle \mbox{ }\mbox{ }\mbox{ }\mbox{ }\mbox{ }\mbox{ }\mbox{ }\mbox{ }\mbox{ }\mbox{ }\mbox{ } S_- |\eta\rangle = \eta |\eta\rangle\mbox{ }\mbox{ }\mbox{ }\mbox{ } \nonumber
\end{eqnarray}
The center of mass or external degree of freedom can be represented in either the position or momentum basis. In the coherent state basis the Hamiltonian Eq.~(\ref{hamiltonian}) can be written in its Q-representation \cite{Wei, WM, Scu, CPP} as [cf Eq.~(2.8) of \cite{ABH}]
\begin{eqnarray}
\label{qrep}H(\{\zb_\bk\}, \{z_\bk\}, \etab, \eta, \bX) = \frac{M \dot{\bX}^2}{2} +\hbar \w_o \etab\eta +\hbar\sum_\bk [\w_\bk \zb_\bk z_\bk + \etab {\rg}_\bk(\bX) z_\bk +\zb_\bk \gb_\bk(\bX) \eta].
\end{eqnarray}
The transition matrix elements between the initial and final coherent states are then
\begin{equation}
K(t,0) = \langle \{\zb_{f\bk}\} \etab_f \bX_f,t |\exp(-\frac{{\rmi}}{\hbar} H t) | \{z_{i\bk}\} \eta_i \bX_i,0 \rangle.
\end{equation}

Using the completeness property of the (EMF and Grassmann) coherent state basis to facilitate time-discretization of the transition matrix \cite{FeyVer} puts the transition matrix elements in a coherent state path integral representation. After inserting the Q-representation, the transition elements transform into a sum over paths in the coherent state labels. Having done the above the transition matrix becomes a triple functional integral:
\begin{eqnarray}
\label{transamp1}K(t,0) = \int D\bX \int D\etab D\eta \prod_\bk D\zb_\bk Dz_\bk \exp[\etab_f \eta(t) &+&\sum_\bk \zb_{f\bk} z_\bk(t) -\frac{{\rmi} M}{2} \int_0^t \dot{\bX}\mbox{ } {\rmd} s] \mbox{ }\rme^{{\rmi} \w_o t/2} \nonumber\\
\times \exp\bigg[-\int_0^t \big(\etab \dot{\eta} + {\rmi}\w_o \etab \eta &+& \sum_\bk \zb_\bk \dot{z}_\bk +{\rmi}\sum_\bk \w_\bk \zb_\bk z_\bk {\rmi}\sum_\bk \etab {\rg}_\bk(\bX) z_\bk +{\rmi}\sum_\bk \zb_\bk \gb_\bk(\bX) \eta \big) {\rmd} s \bigg]
\end{eqnarray}
In this form the transition matrix elements can be evaluated exactly by a combination of stationary phase and correlation function methods which exploit the truncating properties of Grassmann variables. The order of evaluation will be the EMF, COM, and then Grassmann functional integrals. The details follow.

\subsection{EMF Path Integral}
First, the EMF coherent state part of the triple path integral can be evaluated by the stationary phase method \cite{FeyVer}. The variational equations of motion for the electromagnetic  field variables in Eq.~(\ref{transamp1}) are
\begin{equation}
\dot{z}_\bk = -{\rmi} \w_\bk z_\bk -{\rmi} \gb_\bk(\bX) \eta
\end{equation}
which have integral solutions [cf Eq.~(2.14) of \cite{AH}]
\begin{equation}
z_\bk(s) = z_{i\bk} e^{-{\rmi}\w_\bk s} -{\rmi} \int_0^s {\rmd} r \mbox{ } \gb_\bk(\bX(r)) e^{-{\rmi} \w_\bk (s-r)} \eta(r).
\end{equation}
The transition amplitude from an initial EMF vacuum ($\{z_{i\bk}\}=0$) to an arbitrary final state becomes
\begin{eqnarray}
\label{transamp2}K(t,0) &=& \int D\bX \int D\etab D\eta \exp\bigg[\etab_f \eta(t) -\int_0^t \big(\etab \dot{\eta} + {\rmi}\w_o \etab \eta +\frac{{\rmi} M}{2}\dot{\bX}\mbox{ }) {\rmd} s \bigg] \rme^{{\rmi} \w_o t/2} \nonumber\\
&\times&\prod_\bk \exp\bigg[-{\rmi}\int_0^t {\rmd} s \gb_\bk(\bX(s)) e^{-{\rmi}\w_\bk (t-s)} \zb_{f\bk}\mbox{ }\eta(s) -\int_0^t {\rmd} s \int_0^s {\rmd} r {\rg}_\bk(\bX(s)) \gb_\bk(\bX(r)) e^{-{\rmi}\w_\bk (s-r)} \etab(s)\eta(r)\bigg].
\end{eqnarray}
The path integral for the EMF degrees of freedom is now complete.

\subsection{COM Path Integral}
Second, the position path integral can be evaluated as a set of 0, 1 and 2 point functions. Note in the transition amplitude of Eq.~(\ref{transamp2}) that since the EMF is taken to be in an initial vacuum, any source term for $\eta(s)$ will be proportional to $\eta_i$. The variational equation of motion derived from Eq.~(\ref{transamp2}) for $\eta(s)$ yields
\begin{eqnarray}
\dot{\eta}(s) = -{\rmi}\w_o\eta(s) -{\rmi}\int_0^s {\rmd} r \sum_\bk {\rg}_\bk(\bX(s))\gb_\bk(\bX(r))\rme^{-{\rmi}\w_\bk(s-r)}  \eta(r)
\end{eqnarray}
with the boundary condition $\eta(0) = \eta_i$. Therefore $\eta(s) = u(s) \eta_i$.  We use  this to expand the exponent in the transition amplitude of Eq.~(\ref{transamp2}). Due to the nilpotency of the Grassmann variables (i.e.\ $\eta_i^2 =0$) it will truncate after the first term in the expansion.

After expanding and truncating the integrand, the position path integral is
\begin{eqnarray} \int D\bX \exp\bigg[&-&\frac{{\rmi} M}{2}\int_0^t{\rmd} s\mbox{ }\dot{\bX} \bigg] \\
\times\bigg[1 &-&{\rmi}\int_0^t {\rmd} s \sum_\bk \gb_\bk(\bX(s)) e^{-{\rmi}\w_\bk (t-s)} \zb_{f\bk}\mbox{ }\eta(s) -\int_0^t {\rmd} s \int_0^s {\rmd} r \sum_\bk {\rg}_\bk(\bX(s)) \gb_\bk(\bX(r)) e^{-{\rmi}\w_\bk (s-r)} \etab(s)\eta(r)\bigg]. \nonumber
\end{eqnarray}

There are thus three correlation functions which need to be computed. First the spatial mode functions must be chosen in order to specify the targeted correlation functions. For an electromagnetic field in free space (no cavity or boundaries)
\begin{equation}
{\rg}_\bk(\bX) = \frac{\la}{\sqrt{\w_k}} \exp({\rmi} \bk \cdot \bX),
\end{equation}
the correlations functions are computed in Appendix A. Substituting these expressions back into the Eq.~(\ref{transamp2}) gives for the transition amplitude
\begin{eqnarray}
\label{transamp3}K(t,0) = \int D\etab D\eta \exp\bigg[&+&\frac{{\rmi} M}{2 t} (\bX_f -\bX_i)^2 +\etab_f \eta(t) -\int_0^t \big(\etab \dot{\eta} +{\rmi}\w_o \etab \eta \big) {\rmd} s \bigg] \bigg(\frac{M}{2\pi{\rmi} t}\bigg)^{3/2} \rme^{{\rmi} \w_o t/2} e^{-{\rmi}\w_\bk (s-r)} \nonumber\\
\times\bigg\{1 &-&{\rmi}\int_0^t {\rmd} s \sum_\bk \frac{\la}{\sqrt{\w_\bk}} \exp\bigg[{\rmi}\frac{s}{t}\bk\cdot(\bX_f-\bX_i) -\frac{{\rmi}}{2M} \frac{s(t-s)}{t} \bk^2 \bigg] e^{-{\rmi}\w_\bk (t-s)} \zb_{f\bk}\mbox{ }\eta(s) \\
&-&\int_0^t {\rmd} s \int_0^s {\rmd} r \sum_\bk \frac{\la^2}{\w_\bk} \exp\bigg[-{\rmi}\frac{s-r}{t}\bk\cdot(\bX_f-\bX_i) -\frac{{\rmi}}{2M}\frac{(t-(s-r))(s-r)}{t}\bk^2\bigg] \etab(s)\eta(r)\bigg\}. \nonumber
\end{eqnarray}
The path integral for the external degrees of freedom is now complete.

\subsection{Qubit Path Integral}
Finally, the Grassmann variable path integral can be evaluated along its stationary path. The variational equation of motion for the Grassmann field variable in Eq.~(\ref{transamp3}) is
\begin{eqnarray}
\dot{\eta}_t(s) =-{\rmi}\w_o\eta_t(s) -\int_0^s {\rmd} r\sum_\bk \frac{\la^2}{\w_\bk} \mbox{ }\mu_t(s-r)\mbox{ } \eta_t(r)
\end{eqnarray}
with the definition:
\begin{equation}
\mu_t(s) = \exp\bigg[-{\rmi}\w_\bk s -{\rmi}\frac{s}{t}\bk\cdot(\bX_f-\bX_i) -\frac{{\rmi}}{2M}\frac{s(t-s)}{t}\bk^2\bigg].
\end{equation}
Note that the final time $t$ enters as a parameter in the variational equation of motion just as the mass or position do. The reason for this is that the above variational equation of motion is for the evolution of the atom from an initial time to a final time, so the time is an explicit parameter.

Rewriting the above variational equation in Laplace space allows the non-local integral part to be transformed with the convolution theorem. The solution is in terms of an inverse Laplace transform,
\begin{equation}
\label{soln1}\eta_t(s) = \eta_i u_t(s) = \frac{\eta_i}{2\pi{\rmi}}\int_{\gamma-{\rmi}\infty}^{\gamma+{\rmi}\infty} \frac{e^{s z} {\rmd} z}{z +{\rmi}\w_o +\tilde{\mu}(z)}
\end{equation}
with the definition:
\begin{equation}
\tilde{\mu_t}(z) = \frac{\la^2}{\w_\bk}\int_0^\infty \rme^{-sz}\exp\bigg[-{\rmi}\w_\bk s -{\rmi}\frac{s}{t}\bk\cdot(\bX_f-\bX_i) -\frac{{\rmi}}{2M}\frac{s(t-s)}{t}\bk^2\bigg] {\rmd} z.
\end{equation}
The solution thus becomes a contour integral. The pole of the denominator in Eq.~(\ref{soln1}) can be found to $O(\la^2)$
\begin{equation}
z_o = -{\rmi}\w_o -\tilde{\mu}(-{\rmi}\w_o) +O(\la^4).
\end{equation}
Finding the pole to order $O(\la^2)$ gives a solution to the same order:
\begin{eqnarray}
\label{ufunction} u_t(s) = \rme^{-{\rmi}\w_o t} \exp\bigg\{ -\la^2 t \sum_\bk \frac{1}{\w_\bk}\int_0^\infty {\rmd} s \exp\bigg[-{\rmi}\bigg(\w_\bk -\w_o +\frac{\bk\cdot(\bX_f -\bX_i)}{t} -\frac{\bk^2}{2M}\bigg)s -\frac{{\rmi} \bk^2}{2Mt}s^2\bigg] \bigg\}.
\end{eqnarray}

Evaluating the transition amplitude along its stationary path with the second order pole approximation yields an expression for the transition matrix that is second order in its action:
\begin{eqnarray}
\label{transamp4}K(t,0) = \bigg(\frac{M}{2\pi{\rmi} t}\bigg)^{3/2} \exp\bigg[{\rmi} \w_o t/2 &+&\frac{{\rmi} M}{2 t} (\bX_f -\bX_i)^2 +O(\la^4) \bigg] \\
\exp\bigg[\etab_f \eta_t(t) &-&{\rmi}\int_0^t {\rmd} s \sum_\bk \frac{\la}{\sqrt{\w_\bk}} \exp\big[-{\rmi}\w_\bk (t-s) +{\rmi}\frac{s}{t}\bk\cdot(\bX_f-\bX_i) -\frac{{\rmi} s(t-s)}{2Mt}\bk^2\big] \zb_{f\bk}\mbox{ }\eta_t(s)\bigg]. \nonumber
\end{eqnarray}
All three functional integrals are now evaluated. In the next section we proceed to derive the evolutionary operator for the density matrix by combining the transition amplitudes into a closed loop.

\section{Evolutionary Operator}
At this point the expression of Eq.~(\ref{transamp4}) for the transition amplitude can be combined with its counterpart propagating backwards in time and traced over all final EMF states. The result gives the evolutionary operator for the reduced density matrix (we may call it the reduced propagator, for short),
\begin{eqnarray}
\label{reducedprop} J_R = \int \rmd\bX_f \prod_\bk \rmd\mu(z_{f\bk}) K(t, 0) K^*(t, 0),
\end{eqnarray}
and is formed by integrating out the environmental variables which in our case are the EMF and the atom's motional degrees of freedom.

The evolution of the qubit density matrix elements with back-action from the EMF and the atomic motion can be calculated from the reduced propagator
\begin{eqnarray}
\label{evolved1}\rho_R(t) = \int {\rmd}\mu(\eta_i){\rmd}\mu(\etap_i){\rmd}\mu(\bX_i) \mbox{ }J_R \mbox{ }\rho_A(0)\otimes\rho_\bX(0).
\end{eqnarray}
The functions $\rho_A(0)$ and $\rho_\bX(0)$ are initial states for the 2-level atomic and external degrees of freedom, respectively:
\begin{eqnarray}
\rho_A(t)=\rho_{00}(0) + \etab_i \rho_{10}(0) + \etap_i \rho_{01}(0) + \etab_i \etap_i \rho_{11}(0)
\end{eqnarray}
\begin{eqnarray}
\rho_\bX(t)=\Phi(\bX_i) \Phi^*(\bX_i)
\end{eqnarray}
The function $\Phi(\bX)$ is the initial (external) center of mass wavefunction of the atom. From Eq.~(\ref{evolved1}) the on and off-diagonal components of the reduced density matrix elements evolved to time $t$ are given by
\begin{equation}
\rho_{11}(t) = \rho_{11}(0) \bigg(\frac{M}{2\pi{\rmi} t}\bigg)^{3} \int{\rmd}\bX_f \int{\rmd}\bXp_i \int{\rmd}\bX_i \Phi(\bX_i) \Phi^*(\bXp_i) \exp\Bigg\{ \frac{{\rmi} M}{2 t} (\bX_f -\bX_i)^2 -\frac{{\rmi} M}{2 t} (\bX_f -\bXp_i)^2 \bigg\} {\rm \bar{u}_t}(t) {\rm u_t}(t)
\end{equation}
\begin{equation}
\rho_{10}(t) = \rho_{10}(0) \bigg(\frac{M}{2\pi{\rmi} t}\bigg)^{3} \int{\rmd}\bX_f \int{\rmd}\bXp_i \int{\rmd}\bX_i \Phi(\bX_i) \Phi^*(\bXp_i) \exp\Bigg\{ \frac{{\rmi} M}{2 t} (\bX_f -\bX_i)^2 -\frac{{\rmi} M}{2 t} (\bX_f -\bXp_i)^2 \bigg\} {\rm u_t}(t).
\end{equation}
The EMF, as previously stated, is in a vacuum state, but the choice of an initial center of mass wavefunction has not yet been made. To closely model an atom with fixed position and momentum, we use a minimum uncertainty Gaussian wavefunction centered at $(\bX_o=0,\bP_o=0)$.
\begin{eqnarray}
\Phi(\bX) = \pi^{-3/4} \sigma^{-3/2} \exp\bigg[-\frac{\bX^2}{2\sigma^2}\bigg]
\end{eqnarray}
Such an initial wavefunction simplifies the expressions for the diagonal and off-diagonal matrix elements of the qubit.

The result for the off-diagonal components which measures the coherence of the qubit under such conditions is shown here:
\begin{eqnarray}
\rho_{10}(t) = \rho_{10}(0) \frac{4}{\sqrt{\pi}} \bigg( \frac{M^2\sigma^2}{t^2 -2{\rmi} M\sigma^2 t}\bigg)^{3/2} \int_0^\infty {\rmd} x \mbox{ }x^2 u(x,t) \exp\bigg[-\frac{M^2\sigma^2}{t^2 -2{\rmi} M\sigma^2 t} x^2 \bigg]
\end{eqnarray}
The function $u(x,t)$ is given by Eq.~(\ref{ufunction}) with $x=|\bX_f -\bX_i|$.

The evolution of the coherence function is found to follow an exponential decay with a decay rate slightly faster than in the infinite mass case. The percentage change in the decoherence rate of the off-diagonal versus the the stationary qubit case is plotted in Fig.~(1). The decay rate increases with decreasing mass and matches the stationary qubit result given by \cite{ABH} in the limit of infinite mass. We expect that a qubit in a smaller mass object is more affected by recoil than a qubit in heavy mass. The variation in the decoherence rate with changes in the external wavefunction size is relatively flat and cannot reliably be resolved with the available computing power and machine accuracy. We find that so long as the resonant frequency is small enough or the mass large enough that the atomic recoil velocity is non-relativistic, which is where this theory is valid, then the motional decoherence will contribute negligibly to the decay of the qubit.
\begin{figure}
\includegraphics{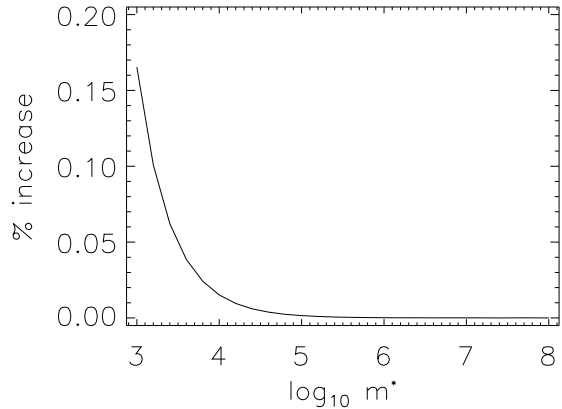}
\caption{\label{fig1} A plot of the percentage increase in the
decoherence of the off-diagonal matrix elements of the reduced
density matrix versus the non-dimensionalized mass ($m^*=\frac{M
c^2}{\hbar\omega_o}$). The decoherence rate increases as the mass
of the atom containing the qubit is decreased. As the mass is
increased the decoherence rate asymptotes to the value of a
stationary atom obtained by Anastopoulos and Hu \cite{ABH}. This
is consistent with a smaller mass qubit being more affected by its
recoil than a heavy mass qubit. Typical experimental parameters
fall to the right end of the shown plot.}
\end{figure}

\section{Discussion}
Often, one may separate the dynamics of an atom's motion from those of its internal degrees of freedom by arguing that the time scales associated with the motion of the atom are much longer than those of the two level activity. This is the rationale behind the adiabatic approximation adopted for most considerations of the atomic dynamics. However, coherence requirements in quantum computing implementations may prompt one to question this assumption. One aim of our investigation is to test for non-adiabatic effects in atomic quantum computing schemes. Another is to describe the effect of recoil from the emission and re-absorption of virtual particles in the atom-EMF interaction upon the center of mass motion. These two problems correspond to the two aspects described in the Introduction. Here we consider the second aspect mentioned above, aiming at the effect of quantum motional decoherence of the qubit, i.e., the back-action of atomic motion on a two level system in free space as mediated by the EMF.

We find that the inclusion of the external degrees of freedom only slightly alters the decoherence and relaxation rates as compared to a stationary atom. Typical experimental parameters fall to the right end in the plot of Fig.~(1). A Rubidium atom used as a qubit would have a non-dimensionalized mass of approximately ${\rm Log} (\frac{M c^2}{\hbar \w_o}) =8$, which places it in a regime in which the effect of motion-induced decoherence is negligible. For optical qubit transition frequencies in general, motion-induced decoherence will not be a factor unless the mass of the qubit is four to five orders of magnitude smaller than the mass of a typical alkali atom. One can conclude tentatively that in general AMO implementations, motion-induced decoherence of a free qubit is negligibly small. Since the calculation done here is coherent and non-Markovian, one can view our result as confirming the validity of the adiabatic approximation in alkali atom qubits.

Although the result of the calculation is the expected one, the technique described here is the first able to compute the decoherence of a qubit coupled to its own quantized COM without any form of Markovian approximation, while allowing the qubit-EMF coupling to be the non-linear form derived from the EMF spatial mode functions. Useful applications of this method will include any situations in which the COM motion of an atom back-acts onto its internal qubit dynamics and the full multi-mode structure of the EMF is relevant. Two such examples, as drawn from the references cited in the Introduction, are an atomic qubit in a cavity and an atom with entangled qubit and EMF degrees of freedom. In the former, the presence of the cavity walls increases the cavity mode recoils on the atom \cite{ShoreMeystreStenholm91}. The latter is at the center of certain two qubit gate implementations \cite{coldcoll}, with the question there being how well coherence is maintained when a qubit is entangled both internally and externally. Calculation in that case can provide an important feasibility test of quantum computing applications which utilize such entanglement.\\

{\bf Acknowledgements} This work is supported in part by NSF grant PHY98-00967 and by a contract under ARDA.

\appendix
\section{COM Functional Integral}
The position path integral which needs to be evaluated is:
\begin{eqnarray} \int D\bX \exp\bigg[&-&\frac{{\rmi} M}{2}\int_0^t{\rmd} s\mbox{ }\dot{\bX} \bigg] \\
\times\exp\bigg[&-&{\rmi}\int_0^t {\rmd} s \sum_\bk \gb_\bk(\bX(s)) e^{-{\rmi}\w_\bk (t-s)} \zb_{f\bk}\mbox{ }\eta(s) -\int_0^t {\rmd} s \int_0^s {\rmd} r \sum_\bk {\rg}_\bk(\bX(s)) \gb_\bk(\bX(r)) e^{-{\rmi}\w_\bk (s-r)} \etab(s)\eta(r)\bigg] \nonumber
\end{eqnarray}
which can be expanded and truncated to:
\begin{eqnarray} \int D\bX \exp\bigg[&-&\frac{{\rmi} M}{2}\int_0^t{\rmd} s\mbox{ }\dot{\bX} \bigg] \\
\times\bigg[1 &-&{\rmi}\int_0^t {\rmd} s \sum_\bk \gb_\bk(\bX(s)) e^{-{\rmi}\w_\bk (t-s)} \zb_{f\bk}\mbox{ }\eta(s) -\int_0^t {\rmd} s \int_0^s {\rmd} r \sum_\bk {\rg}_\bk(\bX(s)) \gb_\bk(\bX(r)) e^{-{\rmi}\w_\bk (s-r)} \etab(s)\eta(r)\bigg] \nonumber
\end{eqnarray}

There are thus three correlation functions which need to be computed. First the spatial mode functions must be chosen in order to specify the targeted correlation functions. For an electromagnetic field in free space (no cavity or boundaries)
\begin{equation}
{\rg}_\bk(\bX) = \frac{\la}{\sqrt{\w_k}} \exp({\rmi} \bk \cdot \bX),
\end{equation}
the correlations functions are:
\begin{equation}
\int_{\bX_i,0}^{\bX_f,t} {\rm D}\bX \exp[-\frac{{\rmi} M}{2}\int_0^t{\rmd} \tau\mbox{ }\dot{\bX}(\tau) ] = \bigg(\frac{M}{2\pi{\rmi} t}\bigg)^{3/2} \exp\bigg[\frac{{\rmi} M}{2 t} (\bX_f -\bX_i)^2\bigg]
\end{equation}
\begin{equation}
\int_{\bX_i,0}^{\bX_f,t} {\rm D}\bX \exp[{\rmi}\bk\cdot\bX(s) -\frac{{\rmi} M}{2}\int_0^t{\rmd} \tau\mbox{ }\dot{\bX}(\tau) ] = \bigg(\frac{M}{2\pi{\rmi} t}\bigg)^{3/2} \exp\bigg[\frac{{\rmi} M}{2 t} (\bX_f -\bX_i)^2 +{\rmi}\frac{s}{t}\bk\cdot(\bX_f-\bX_i) -\frac{{\rmi}}{2M}\frac{s(t-s)}{t}\bk^2\bigg] \\
\end{equation}
\begin{eqnarray} \int_{\bX_i,0}^{\bX_f,t} {\rm D}\bX \exp[-{\rmi}\bk\cdot\bX(s) &+&{\rmi}\bk\cdot\bX(r) -\frac{{\rmi} M}{2}\int_0^t{\rmd} \tau\mbox{ }\dot{\bX}(\tau) ] \\
&=& \bigg(\frac{M}{2\pi{\rmi} t}\bigg)^{3/2} \exp\bigg[\frac{{\rmi} M}{2 t} (\bX_f -\bX_i)^2 -{\rmi}\frac{s-r}{t}\bk\cdot(\bX_f-\bX_i) -\frac{{\rmi}}{2M}\frac{(t-(s-r))(s-r)}{t}\bk^2\bigg] \nonumber
\end{eqnarray}

Substituting these expressions back into the Eq.~(\ref{transamp2}) gives for the transition amplitude:
\begin{eqnarray}
K(t,0) = \int D\etab D\eta \exp\bigg[&+&\etab_f \eta(t) -\int_0^t \big(\etab \dot{\eta} +{\rmi}\w_o \etab \eta \big) {\rmd} s \bigg] \rme^{{\rmi} \w_o t/2} \bigg(\frac{M}{2\pi{\rmi} t}\bigg)^{3/2} \exp\bigg[\frac{{\rmi} M}{2 t} (\bX_f -\bX_i)^2\bigg]e^{-{\rmi}\w_\bk (s-r)} \nonumber\\
\times\bigg[1 &-&{\rmi}\int_0^t {\rmd} s \sum_\bk \frac{\la}{\sqrt{\w_\bk}} \exp\bigg[{\rmi}\frac{s}{t}\bk\cdot(\bX_f-\bX_i) -\frac{{\rmi}}{2M}\frac{s(t-s)}{t}\bk^2\bigg] e^{-{\rmi}\w_\bk (t-s)} \zb_{f\bk}\mbox{ }\eta(s) \\
&-&\int_0^t {\rmd} s \int_0^s {\rmd} r \sum_\bk \frac{\la^2}{\w_\bk} \exp\bigg[-{\rmi}\frac{s-r}{t}\bk\cdot(\bX_f-\bX_i) -\frac{{\rmi}}{2M}\frac{(t-(s-r))(s-r)}{t}\bk^2\bigg] \etab(s)\eta(r)\bigg] \nonumber
\end{eqnarray}\\


\begin{thebibliography}{9}
\bibitem{CCT1}
C. Cohen-Tannoudji, {\it Atomic Motion in Laser Light} eds.\ J. Dalibard, J. M. Ramond, and J. Zinn-Justin {\bf Les Houches, Session LIII, 1990} (Elsevier Science 1991).

\bibitem{CCTandWDP}
C. Cohen-Tannoudji and W. D. Phillips, Physics Today {\bf 43 10}, 33 (1990).

\bibitem{phillips98}
W. D. Phillips, Rev.\ Mod.\ Phys.\ {\bf 70}, 21 (1998).

\bibitem{CCT2}
C. Cohen-Tannoudji, Rev.\ Mod.\ Phys.\ {\bf 70}, 707 (1998).

\bibitem{calarco}
T. Calarco, H. -J. Briegel, D. Jaksch, J. I. Cirac, and P. Zoller, Fortschr.\ Phys.\ {\bf 48 9-11}, 945 (2000).

\bibitem{brennen}
G. K. Brennen, C. M. Caves, P. S. Jessen, and I. H. Deutsch, Phys.\
Rev.\ Lett.\ {\bf 82}, 1060 (1999).

\bibitem{deutsch}
I. H. Deutsch and G. K. Brennen, Fortschr.\ Phys.\ {\bf 48 9-11}, 925 (2000).

\bibitem{milburn}
G. J. Milburn, Fortschr.\ Phys.\ {\bf 48 9-11}, 957 (2000).

\bibitem{coldcoll}
D. Jaksch, H. -J. Briegel, J. I. Cirac, C. W. Gardiner, and P. Zoller, Phys.\ Rev.\ Lett.\ {\bf 82}, 1975 (1999).

\bibitem{poyatos}
J. F. Poyatos, J. I. Cirac, and P. Zoller, Fortschr.\ Phys.\ {\bf 48 9-11}, 785 (2000).

\bibitem{DubetskyBerman96}
B. Dubetsky and P. R. Berman, Phys.\ Rev.\ A {\bf 53}, 390 (1996).

\bibitem{WilkensMeystre92}
M. Wilkens, Z. Bialynicka-Birula, and P. Meystre, Phys.\ Rev.\ A {\bf 45}, 477 (1992).

\bibitem{RenCarmichael95}
W. Ren and H. J. Carmichael, Phys.\ Rev.\ A {\bf 51}, 752 (1995).

\bibitem{VernooyKimble97}
D. W. Vernooy and H. J. Kimble, Phys.\ Rev.\ A {\bf 56}, 4287 (1997).

\bibitem{GrahamWallsZoller92}
Robert Graham, Daniel F. Walls, and Peter Zoller, Phys.\ Rev.\ A {\bf 45}, 5018 (1992).

\bibitem{AH}
A. Anderson and J. J. Halliwell, Phys.\ Rev.\ D {\bf 48}, 2753 (1993); C. Anastopoulos and J. J.  Halliwell, Phys. Rev.\ D {\bf 51}, 6870 (1995); C. Anastopoulos, Phys.\ Rev.\ E {\bf 53}, 4711 (1996).

\bibitem{FeyVer}  R. P. Feynman and A. R. Hibbs, {\it Quantum Mechanics and Path Integrals} (McGraw-Hill, New York, 1965) ; R. P. Feynman and F. L. Vernon, Ann.\ Phys.\ {\bf 24}, 118 (1963).

\bibitem{ctp}
J. Schwinger, J. Math.\ Phys.\ {\bf 2} (1961) 407; L. V. Keldysh,
Zh.\ Eksp.\ Teor.\ Fiz.\ {\bf 47 }, 1515 (1964) [Engl.\ trans.\
Sov.\ Phys.\ JEPT {\bf 20}, 1018 (1965)]. G. Zhou, Z. B. Su, B.
Hao and L. Yu, Phys.\ Rep.\ {\bf 118}, 1 (1985); Z. B. Su, L. Y.
Chen, X. T. Yu and K. C. Chou, Phys.\ Rev.\ {\bf B37}, 9810
(1988). E. Calzetta and B. L. Hu, Phys.\ Rev.\ {\bf D40}, 656
(1989).

\bibitem{ABH}
C. Anastopoulos and B. L. Hu, Phys.\ Rev.\ A {\bf 62} 033821 (2000).

\bibitem{ohnuki}
Ohnuki Y, Kashiwa T 1978 Coherent states of Fermi operators and the path integral {\it Coherent states: applications in physics and mathematical physics} eds J Klauder and B Skagerstam (Singapore: World Scientific) 449-465.

\bibitem{perelomov}
A. Perelomov, {\it Generalized coherent states and their applications} (Berlin: Springer 1986).

\bibitem{WM}
D. F. Walls and G. J. Milburn, {\it Quantum Optics} ( Springer Verlag, Berlin, Heidelberg, 1994).

\bibitem{Scu}
O. Scully and M. Suhail Zubairy, {\it Quantum Optics} (Cambridge University Press, Cambridge,  1997).

\bibitem{Wei}
M. Weissbluth, {\it Photon-Atom Interactions} (Academic Press, San Diego, 1988).

\bibitem{CPP}
G. Compagno, R. Passante and F. Persico, {\it Atom-Field Interactions and Dressed Atom} (Cambridge University, Cambridge, 1995).

\bibitem{ShoreMeystreStenholm91}
B. W. Shore, P. Meystre, and S. Stenholm, J. Opt.\ Soc.\ Am.\ B {\bf 8}, 903 (1991).

\end{thebibliography}
\end{document}